\documentclass{PoS}

\usepackage{amsmath,amssymb,amsbsy}
\usepackage{exscale,mathrsfs,stmaryrd,eucal}
\usepackage[]{graphicx}
\usepackage{psfrag}

\newcommand{\be}{\begin{equation}}
\newcommand{\ee}{\end{equation}}
\newcommand{\ba}{\begin{eqnarray}}
\newcommand{\ea}{\end{eqnarray}}

\title{Cosmological first-order phase transitions beyond the standard inflationary scenario}

\ShortTitle{1st-order phase transitions beyond the standard inflationary scenario}

\author{\speaker{Aleksandar Raki\'c}\\
        ITPA Universit\"at W\"urzburg, Am Hubland, 97074 W\"urzburg, Germany\\
        E-mail: \email{rakic@astro.uni-wuerzburg.de}}

\author{Dennis Simon\\
        ITPA Universit\"at W\"urzburg, Am Hubland, 97074 W\"urzburg, Germany\\
        E-mail: \email{dsimon@astro.uni-wuerzburg.de}}

\author{Julian Adamek\\
        ITPA Universit\"at W\"urzburg, Am Hubland, 97074 W\"urzburg, Germany\\
        E-mail: \email{jadamek@physik.uni-wuerzburg.de}}

\author{Jens C.~Niemeyer\\
        Institut f\"ur Astrophysik, Universit\"at G\"ottingen, Friedrich-Hund-Platz 1, 37077 G\"ottingen, Germany\\
        E-mail: \email{niemeyer@astro.physik.uni-goettingen.de}}

\abstract{
Motivated by cosmological first-order phase transitions we examine the nucleation and evolution of vacuum bubbles in non-vacuum environments. Non-standard backgrounds can be relevant in the context of rapid tunneling processes on the landscape. Utilising complex time methods, we show that tunneling rates can be notably modified in the case of dynamical FRW backgrounds. We give a classification of the importance of the effect in terms of the relevant dynamical time scales. For both the bubble nucleation and evolution analysis we make use of the thin-wall approximation. From the classical bubble evolution on homogeneous matter backgrounds via the junction method, we find that the inflation of vacuum bubbles is very sensitive to the presence of ambient matter and quantify this statement. We also employ inhomogeneous matter models (LTB) and models that undergo a rapid phase transition (FRW) as a background and discuss in which cases potentially observable imprints on the bubble trajectory can remain.
}

\FullConference{International Workshop on Cosmic Structure and Evolution - Cosmology2009,\\
		September 23-25, 2009\\
		Bielefeld, Germany}

\begin{document}
\newcommand{\rmd}{\rm d}

\section{Motivation}
Given the cosmic no hair conjecture, as well as the very long lifetimes of metastable states on the landscape, cosmological tunneling has been studied mainly between pure vacuum states. However, in the context of the string landscape there are recently proposed scenarios where tunneling can be catalysed. When very rapid tunneling occurs a vacuum bubble can nucleate and find itself on a non-vacuum background that had not enough time to evolve to a de Sitter spacetime. Rapid tunneling occurs for instance in chain inflation \cite{Freese05} where many coupled fields are at work providing a series of catalysed tunneling processes through many minima on the landscape, or in the context of resonance or DBI tunneling \cite{Sarangi07,Tye06}.

Here, we will study the nucleation and the classical evolution of vacuum bubbles on non-vacuum backgrounds in the thin-wall approximation; for a more ample treatment see \cite{Simon09}. A background we utilise for both analyses is a flat FRW model that will serve as the setting for ambient phase transitions, like reheating. For the nucleation part we study the effects of a power law inflating FRW background on the tunneling rates, using a complex time path formalism. For the classical part we probe the effects of homogeneous (FRW) and inhomogeneous (LTB) matter backgrounds on the evolution of the vacuum bubbles by employing Israel's junction method \cite{Israel}. Note that inhomogeneous backgrounds can also be relevant in the context of resonance tunneling, cf.~\cite{Saffin08,Copeland07,Tye09}.

\section{Tunneling Rates}
The semiclassical calculation of tunneling rates between states of pure vacuum differing only in their values for $\Lambda$ is a well known and commonly used result, first carried out by Coleman and De Luccia (CdL) \cite{CDL80}. CdL tunneling produces a bubble of new vacuum whose interior, and this is the main result of the CdL calculation, is also a de Sitter spacetime and thus can be used to model inflation. Hence, spherically symmetric bubbles of new vacuum expand into the old vacuum and so a first-order phase transition (PT) can be realised.

However, the CdL mechanism and its variations rely on the participating states to be in a pure vacuum. Here we want to relax the de Sitter symmetry of the initial state and see how this can affect the tunneling rate. Of course, it is not at all clear how to do a calculation of tunneling rates  on completely arbitrary backgrounds, and so we diverge only little from de Sitter spacetime in that we are taking background Friedmann universes in this section which will affect the tunneling rate through their time-dependent Hubble rate.

Assuming spherical symmetry, a flat FRW background, and neglecting gravitational backreaction, the dynamics of the bubble becomes a $1+1$ dimensional problem with the effective action
\begin{equation}
 S = \int {\rm d} \eta \left[ \frac{4 \pi}{3} \epsilon~a^4 \left(\eta\right) \bar{r}^3 \left(\eta\right) - 4 \pi \sigma~a^3 \left(\eta\right) \bar{r}^2 \left(\eta\right) \sqrt{1 - \left(\partial_\eta \bar{r}\left(\eta\right)\right)^2}\right] \,,
\label{FRWaction}
\end{equation}
where $\eta$ denotes conformal time, $\bar{r}$ the coordinate radius of the shell, $\epsilon$ the latent heat of the vacuum and $\sigma$ the bubble surface tension. The scale factor enters the equations of motion and introduces an explicit time dependence and this will make the tunneling rate time dependent as well.

\begin{figure}[ht]
\psfrag{ImS}[b][b]{\small $\mathrm{Im}S~/~\mathrm{Im}S_\mathrm{dS}$}
\psfrag{epsilon}[t][t]{\small $\left| \partial_t H / H^2 \right|$}
\psfrag{proper}[l][l]{\small proper time average}
\psfrag{conformal}[l][l]{\small conformal time average}
\psfrag{tcross=2tH}[l][l]{\small $t_{\rm cross} \sim 2 t_H$}
\centering
\includegraphics[width=0.45\textwidth]{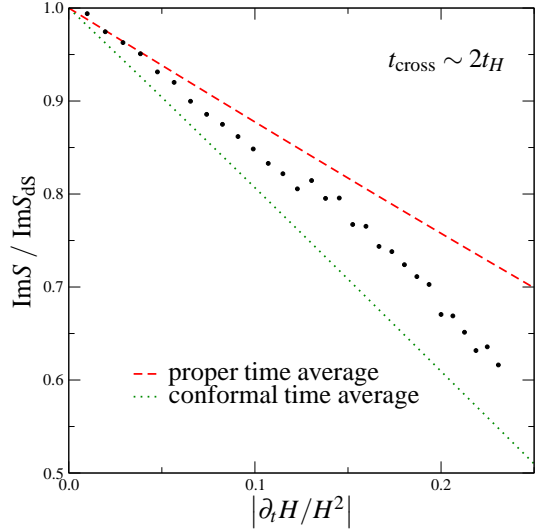}
\caption{\label{figpl}
Considering a FRW background with power law inflation numerical values for the tunneling rate as a function of the slow roll parameter $\partial_t H / H^2$ are given (black circles). The values are normalised to the quasistatic approximation, that is the limit of the result for the tunneling on de Sitter background (2.3) when fixing $H = H\left(\eta_0\right)$ with $\eta_0$ denoting the nucleation time. Three typical time scales are notable here and their comparison can help to categorise the dynamics of the tunneling process. These are the bubble light crossing time $t_{\rm cross}$, the inverse of the background Hubble rate $t_H$ as well as its respective rate of change $t_{\dot{H}}$. Only when the bubble crossing time is not negligible w.r.t.~both of the other two time scales the dynamics of the background becomes significantly important for the tunneling rate and neither the Minkowskian nor the quasistatic limit are good approximations. One can improve the quasistatic estimate by employing a proper time averaged (dashed red line) or conformally time averaged (dotted green line) Hubble rate in (2.3).}
\end{figure}

In \cite{KVK96} a compendium of how one can treat tunneling processes in time dependent setups has been presented. Here we want to apply these techniques to the example of a \emph{power law inflating background}, $a = \left(\eta_1 / \eta\right)^{1 + \alpha}$ with $a(\eta_1)\equiv1$, and thereby analyse the effects of a time dependent background on the tunneling rate. Note that for small deformations $\alpha$ this is just an inflationary slow-roll solution with slow roll parameter $-\partial_t H / H^2 \approx \alpha$. For a power law inflation we get, after some transformations, the following equation of motion from the action
\begin{equation}
 \frac{\epsilon}{\sigma} \left(\frac{\eta_1}{\eta}\right)^{1 + \alpha} \sqrt{\left(\partial_{\bar{r}} \eta\right)^2 - 1} = 2 \frac{\partial_{\bar{r}} \eta}{\bar{r}} - 3 \frac{1 + \alpha}{\eta} - \frac{\partial_{\bar{r}}^2 \eta}{\left(\partial_{\bar{r}} \eta\right)^2 - 1} \,.
\end{equation}
This is an equation for $\eta(\bar{r})$, which is an analytic (in general) complex function of real arguments $\bar{r}$ ranging between $0$ and $\infty$. We can solve it only numerically, taking into account the boundary conditions for the canonical momentum $\bar{p}\left(\eta_0\right) = 0$ as well as $\partial_{\bar{r}}\eta\left(0\right) = 0$. We have carried out a parameter study with regard to the relevant model parameters $\alpha$, $\eta_0 / \eta_1$ and $\epsilon / \sigma$ and found that the following three characteristic time scales can be used to describe the physical behaviour: the time scale $t_H$, given by the inverse Hubble rate $H^{-1}\left(\eta\right)$, on which the background (scale factor) changes, the time scale $t_{\dot{H}}$ on which $H$ itself changes, given by $\left|\partial_t H / H\right|^{-1}$ (higher order derivatives of the expansion rate vanish in power law inflation), as well as the light crossing time for the bubble $t_{\rm cross}$. (A) in case that the crossing time is the smallest scale in the setting, i.e.~$t_{\rm cross} \ll t_H , t_{\dot{H}}$, we find that the Minkowskian result $\mathrm{Im}S_{\mathrm{Mink}} = \frac{27 \pi^2}{4} \frac{\sigma^4}{\epsilon^3}$ for the tunneling rate is a good approximation. Then, if the bubble crossing time scale is not much smaller than the inverse Hubble scale, $t_{\rm cross} \gtrsim t_H$, there are two possibilities. (B) if the rate of change of the Hubble time is still the largest time scale $t_H, t_{\rm cross} \ll t_{\dot{H}}$ we find that our result for the tunneling rate in de Sitter spacetime
\begin{equation}
\mathrm{Im} S_{\mathrm{dS}} = \frac{4 \pi^2 \epsilon}{3 H^4} \sinh^2 \frac{1}{4} \ln\left(1 + \left(3 H \sigma / \epsilon\right)^2\right)
\label{quasi}
\end{equation}
in the quasistatic limit $H = H\left(\eta_0\right)$ is a good approximation. However if finally (C) the crossing time cannot be regarded as small w.r.t.~any other time scale, $t_{\rm cross} \gtrsim t_{\dot{H}}$, then the tunneling process does become susceptible of the background dynamics. In case (C) the changing rate of expansion can have a significant effect on the tunneling rate as our numerical analysis confirms -- in fact the tunneling rate in this case is enhanced w.r.t.~the quasistatic approximation. Trying to exploit the de Sitter result (\ref{quasi}) in order to approximate the non-trivial behaviour of case (C) one should, instead of fixing $H = H\left(\eta_0\right)$, average the expansion rate over one bubble crossing time range prior to tunneling and then insert the averaged Hubble rate into (\ref{quasi}), see Fig.~\ref{figpl}.

Another background also possibly of interest for chain inflation is a radiation dominated FRW universe with cosmological constant. Due to the cosmic no-hair theorem in its far future this universe will be vacuum dominated. However, its early phase where it is radiation dominated is characterised by a rapidly decreasing Hubble rate and so the dynamics may have consequences on the tunneling rate similar to the power law inflationary case. One difference is the existence of a particle horizon. The horizon scale is $\sim t$, as are the two background time scales. Because of causality the bubble crossing time is bounded by the horizon scale and thus it cannot become larger than any of the background time scales. As a consequence the corrections from dynamics should in general be small. Also, in the background spacetime there exists an initial singularity. A numerical analysis shows that the rates for nucleation of bubbles of about horizon scale become sensitive to the evolution of the background scale factor up to the vicinity of the initial singularity. Thus, the rates for such bubbles become sensitive also to the details of reheating and even earlier physics: the classical horizon problem also occurs in the context of vacuum tunneling.

\section{Classical Evolution}
In this section we want to study the (subsequent) classical evolution of a bubble on dynamical backgrounds, cf.~\cite{Fischler07}. Like in the previous section we assume the bubble wall to be thin compared to its size. The spacetimes of the de Sitter bubble and the matter background will be matched along a common spherically symmetric hypersurface with induced metric $ h_{ij} {\rmd}y^i {\rmd}y^j \equiv -{\rmd\tau}^2 + R^2{\rmd\Omega}^2$, stress-energy $S_{ij} = -\sigma h_{ij}$ and the bubble surface tension $\sigma$ known from the preceding section. As the dynamical background spacetime we employ the spherically symmetric and radially inhomogeneous Lema\^ itre-Tolman-Bondi (LTB) solution \cite{Lemaitre}:
\begin{equation} 
  {\rmd}s^2 = -{\rmd}t^2 + \frac{\left(r\partial_r a(t,r) + a(t,r)\right)^2}{1+2E(r)}{\rmd}r^2 + a^2(t,r)r^2{\rmd}\Omega^2 \;.
\end{equation}
The free function $E(r)$ can be interpreted as either a measure of the local spatial curvature or as a measure of the total energy per unit mass of a shell at given radius $r$. Including dust and vacuum energy we get the following Einstein equations
\begin{equation}
  \left(\frac{\partial_t a}{a}\right)^2 - \frac{2E}{a^2r^2} = \frac{2M}{a^3r^3} +\frac{\Lambda}{3} \quad \text{and} \quad 8\pi\rho = \frac{2\partial_r M}{a^2r^2\left(r\partial_ra+a\right)} \;.
\end{equation}
Here, $M(r)$, the total mass within a given shell of radius $r$, is another free function and a third free function (the bang time) arises from the integration of the scale factor $a$. But there is also a remaining gauge freedom in $r$ and, when $\partial_r M >0$, one can rescale $r$ such that the function $M(r)$ becomes $M(r) = \frac{4\pi}{3}A r^3$, where $A$ is a constant.

For our setup the junction conditions yield the following equations of motion in the LTB frame,
\begin{equation}
 \partial_t \bar r = \frac{-(1+2E)\bar r\partial_t a +\sqrt{(1+2E)\left(1+2V\right)Y}}{\left(\bar r\partial_{\bar r} a+a\right)\left(2E-2V\right)} \;, \;
 \partial_t \sigma = \rho\frac{\left(\bar r\partial_{\bar r} a+a\right)\partial_t \bar r} {\sqrt{1+2E-\left(\bar r\partial_{\bar r} a+a\right)^2\left(\partial_t \bar r\right)^2}} \,,
\label{ijs}
\end{equation}
where $\bar r(t)$ denotes the bubble trajectory in LTB coordinates, $Y\equiv(\bar r\partial_t a)^2-2E+2V$ with an effective potential $V =  -\left[\frac{\Lambda_-}{3} + \left(\frac{A}{3a^3\sigma} +\frac{\Lambda_+ -\Lambda_-}{24\pi\sigma} +2\pi\sigma \right)^2\right]\frac{R^2}{2}$. There is a geometrical constraint on the surface tension due to the glueing procedure, $4\pi\sigma < \sqrt{\frac{8\pi A}{3a^3} + \frac{\Lambda_+ -\Lambda_-}{3}}$.

\begin{figure}[ht]
\begin{center}
\begin{tabular}{lr}
\psfrag{r/e}[b][b]{\small $\rho / \epsilon_{\mathrm{vac}}$}
\psfrag{6pGs2/e}[t][t]{\small $6 \pi \sigma^2 / \epsilon_{\mathrm{vac}}$}
\psfrag{collapsing}[b][b]{\small \textbf{contracting}}
\psfrag{expanding}[t][t]{\small \textbf{expanding}}
\psfrag{forbidden}[b][b]{\small \textbf{forbidden}}
\psfrag{smallest possible rL}[b][b]{\footnotesize matter dominated universes possible}
\psfrag{vacuum dominated universes}[t][t]{\footnotesize all universes vacuum dominated}
\includegraphics[width=0.45\textwidth]{sigmabounds.eps}
&
\psfrag{y}[][]{\scriptsize{$\rho(t,\bar r)/A$}}
\psfrag{x}[][]{\scriptsize{$\sqrt{\Lambda_+/3}~t$}}
\includegraphics[width=0.442\textwidth]{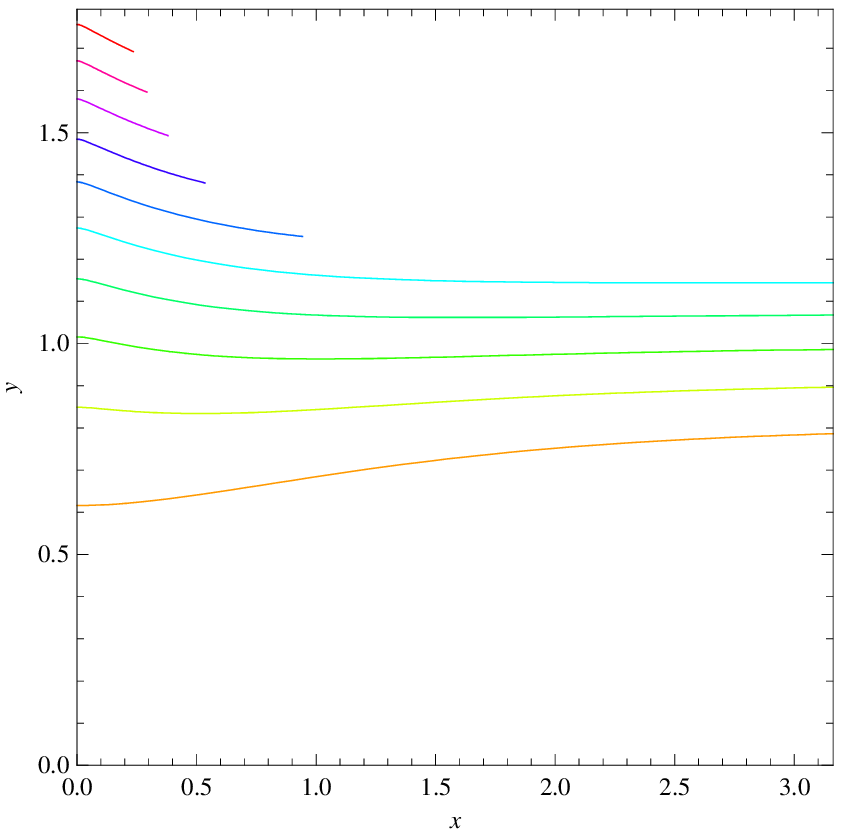}
\end{tabular}
\caption{\label{RhoVsSigm}
\emph{Left}: comovingly nucleated vacuum bubbles on a flat FRW background with dust and cosmological constant (homogeneous limit of LTB). Shown are the possible initial bubble trajectories (as seen by comoving FRW observers) in the parameter space of the governing force balance quantities: the dust density $\rho$, surface tension $\sigma$ and the latent heat of the vacuum $\epsilon_{\mathrm{vac}} \equiv \left(\Lambda_+ - \Lambda_-\right) / 8 \pi$, cf.~(3.4). The expansion of de Sitter bubbles that is generic in the case of a pure vacuum background is here seriously hindered as long as the environment is matter dominated ($\rho > \rho_{\mathrm{vac}} = \Lambda_+ / 8 \pi$). The forbidden area is due to a geometrical constraint from the junction method. \emph{Right}: an LTB model with inhomogeneous initial dust distribution does allow expanding vacuum bubbles. Assuming, say, $\rho_0(r) \propto r^3$ vacuum bubbles that have large nucleation scales are already in a matter dominated environment and are prevented from expanding while smaller nucleated bubbles find themselves in a yet vacuum dominated region and may initially expand.}
\end{center}
\end{figure}

We first consider the \emph{homogeneous limit} of the LTB background, which can be obtained by imposing a homogeneous initial density distribution as well as a vanishing local curvature. The evolution of a comovingly nucleated ($\partial_t \bar{r} = 0$) bubble of new vacuum on such background substantially depends on the pressure balance and so on the relation of surface tension to latent heat and dust density. The initial force balance can be obtained by differentiating (\ref{ijs}),
\begin{equation}
\left. \partial_t^2 \bar{r}\right|_{\partial_t \bar{r} = 0}~=~\frac{1}{a} \left(\frac{\Lambda_+ - \Lambda_-}{24 \pi \sigma} - 2 \pi \sigma - \frac{2 \rho}{3 \sigma}\right) \; .
\label{budget}
\end{equation}
The vital difference to nucleation on vacuum is that the dust obstructs the expansion of the bubble, eventually leading to a contraction. One can understand this from the competing forces: on the one hand the bubble tends to collapse under its surface tension and on the other hand the pressure induced by the latent heat of the vacuum (with $\Lambda_+ > \Lambda_- \geq 0$) works to expand the bubble. On a vacuum dominated background $(\rho < \rho_{\mathrm{vac}} \equiv \Lambda_+ / 8 \pi)$ the pressure force is able to outweigh surface tension. We find however that in a matter dominated background the presence of dust effectively makes the pressure support incapable of upholding the expansion of the bubble, see Fig.~\ref{RhoVsSigm} (left).

\begin{figure}[ht]
\begin{center}
\begin{tabular}{lr}
  \psfrag{Hr}[][]{\scriptsize{$\sqrt{\Lambda_+/3} r$}}
  \psfrag{y}[][]{\scriptsize{$R_\mathrm{cr}\sqrt{k}$}}
  \includegraphics[width=0.437\textwidth]{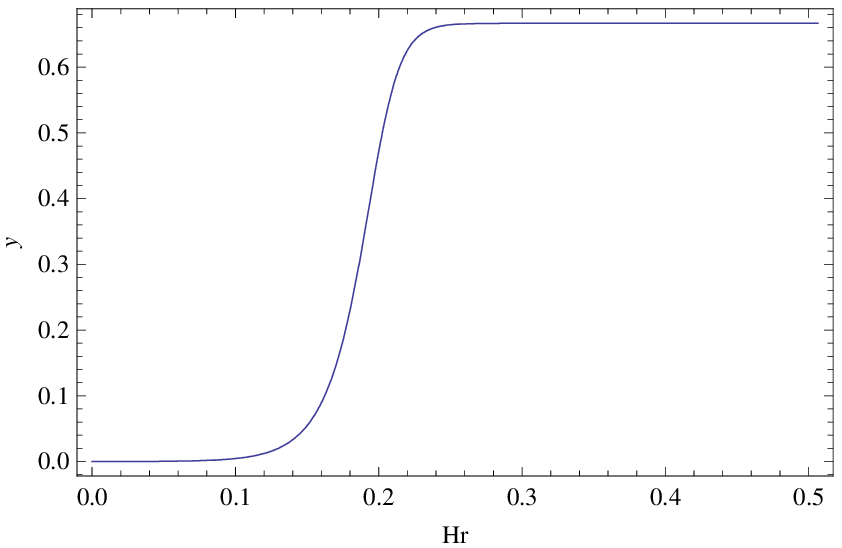}
&
  \psfrag{y}[][]{\scriptsize{$\sqrt{\Lambda_-/3}~\bar r$}}
  \psfrag{x}[][]{\scriptsize{$\sqrt{\Lambda_-/3}~t$}}
  \includegraphics[width=0.437\textwidth]{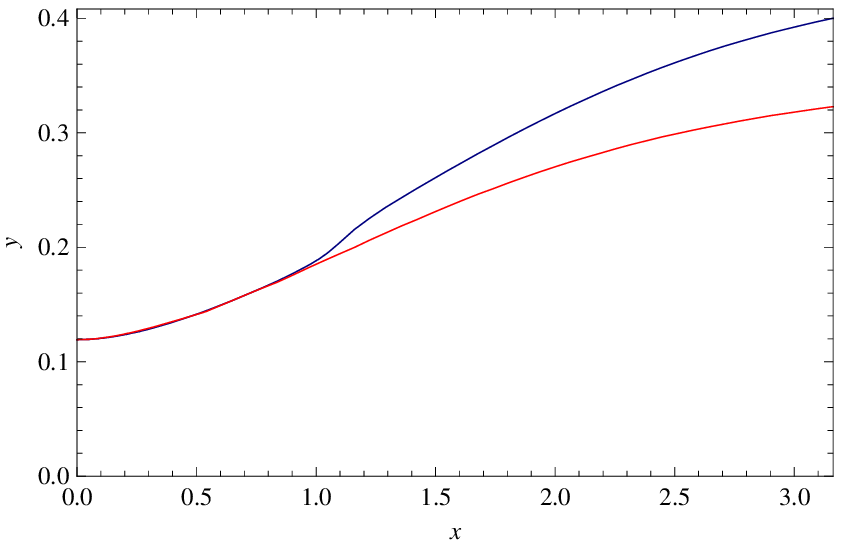}\\

  \psfrag{y}[][]{\scriptsize{$\sqrt{\Lambda_+/3}~\bar r$}}
  \psfrag{x}[][]{\scriptsize{$\sqrt{\Lambda_+/3}~t$}}
  \includegraphics[width=0.437\textwidth]{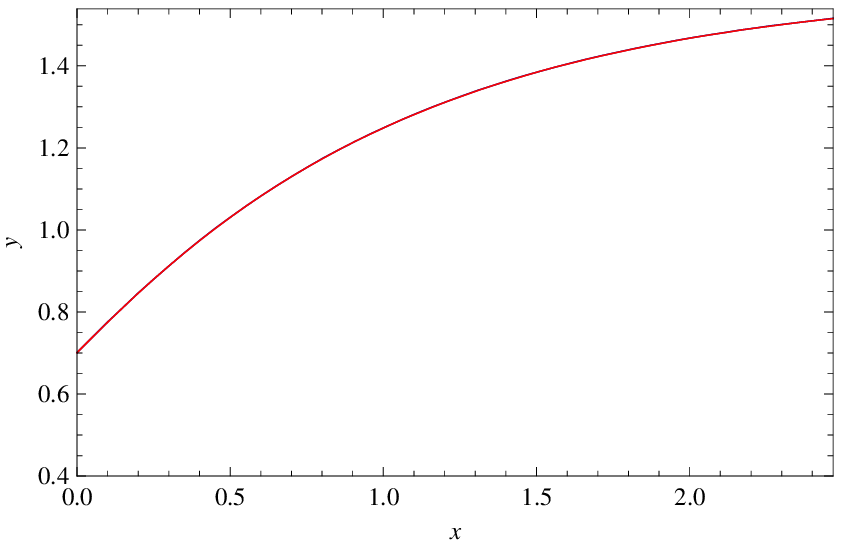}
&
  \psfrag{y}[][]{\scriptsize{$\sigma/\sigma_0$}}
  \psfrag{x}[][]{\scriptsize{$\sqrt{\Lambda_-/3}~t$}}
  \includegraphics[width=0.437\textwidth]{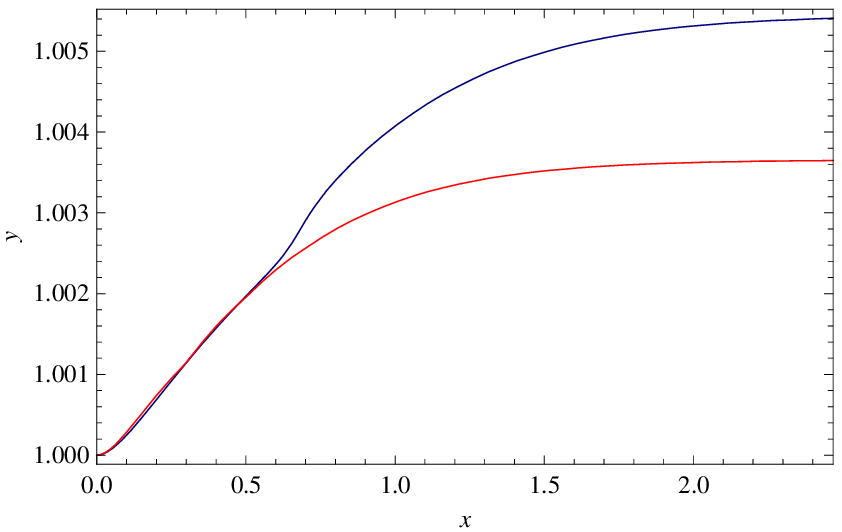}
\end{tabular}
\caption{\label{plot2inhoms}
Numerical evolution of a vacuum bubble on an inhomogeneous matter background (LTB). Expanding bubbles are realised in an inhomogeneous initial dust distribution and are then traversing an additional curvature inhomogeneity of the LTB model - the curvature profile is shown in the \emph{upper left panel}. Comparing bubble trajectories on homogeneous (red) and inhomogeneous (blue) background, a significant difference is only seen in the exterior frame (\emph{upper right}), while in the interior coordinates the transition occurs noteless (\emph{lower left}). The sizeable effect of the transition in the evolution of the bubble surface tension (\emph{lower right}) is to be considered as an artefact of the junction method rather than a physical effect.\vspace{-3mm}}
\end{center}
\end{figure}

Next we will study bubble evolution on the radially \emph{inhomogeneous LTB background}. Our aim is to probe whether the transition of the domain wall through background inhomogeneities can disturb the bubble trajectory. To implement inhomogeneities in the LTB model we can use the local curvature function $E(r)$ as well as the initial density profile $\rho_0(r)\equiv \rho(t_0,r)$. Recalling the results of the homogeneous limit, we see that the bubble would hardly be able to arrive at the inhomogeneities because of the dust in the background. A numerical study indicates that even relaxing the condition of a comoving nucleation, i.e.~considering a bubble with  $\partial_t \bar r(t_0)>0$, does not save the bubble from collapsing whenever there is enough dust in the background. Nevertheless, there exist combinations of $\rho_0(r)$ and $E(r)$ for which the bubble can reach ambient curvature inhomogeneities. For a radially growing initial dust profile $\rho_0(r)$ small nucleated bubbles find themselves in a vacuum dominated region and are able to expand initially, cf.~Fig.~\ref{RhoVsSigm} (right), and can so reach the curvature inhomogeneities. We numerically traced the trajectories of these bubbles: for (potentially physical) interior observers there appears no difference w.r.t.~the corresponding trajectories in the homogeneous limit, cf.~Fig.~\ref{plot2inhoms}. Note, however, that the amplitude of the curvature profile is additionally constrained by demanding the absence of shell crossing singularities.

\begin{figure}[ht]
\begin{center}
\begin{tabular}{lr}
  \psfrag{y}[][]{\scriptsize{$H_0\bar r$}}
  \psfrag{x}[][]{\scriptsize{$H_0 t$}}
  \psfrag{W}[][]{\scriptsize{\fbox{$w=1/3 \rightarrow w=-1$}}}
  \includegraphics[width=0.438\textwidth]{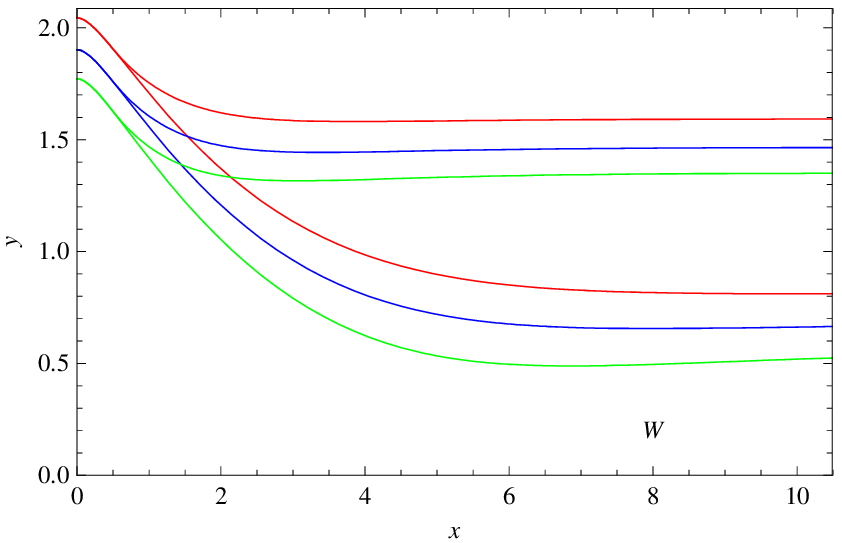}
&
  \psfrag{y}[][]{\scriptsize{$H_0\bar r$}}
  \psfrag{x}[][]{\scriptsize{$H_0 t$}}
  \psfrag{W}[][]{\scriptsize{\fbox{$w=-1 \rightarrow w=1/3$}}}
  \includegraphics[width=0.438\textwidth]{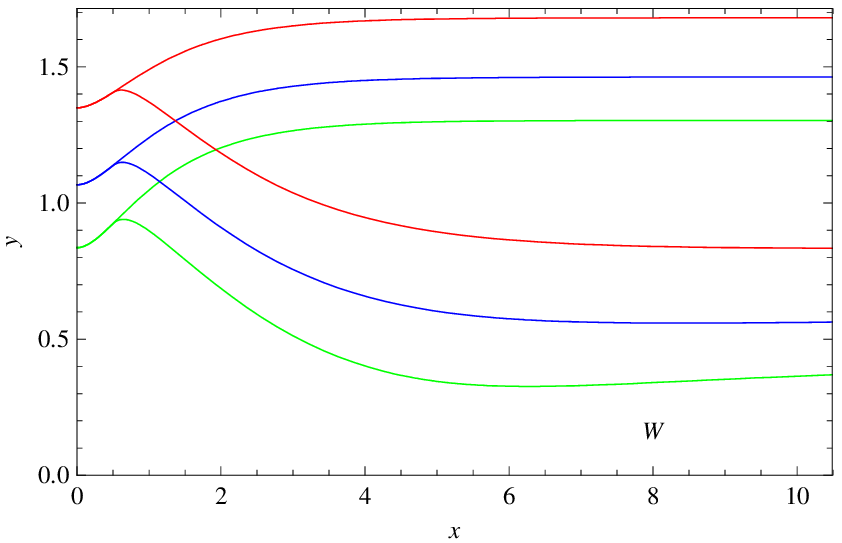}\\

  \psfrag{y}[][]{\scriptsize{$\sqrt{\Lambda_-/3}~\bar r$}}
  \psfrag{x}[][]{\scriptsize{$\sqrt{\Lambda_-/3}~t$}}
  \psfrag{W}[][]{\scriptsize{\fbox{$w=1/3 \rightarrow w=-1$}}}
  \includegraphics[width=0.438\textwidth]{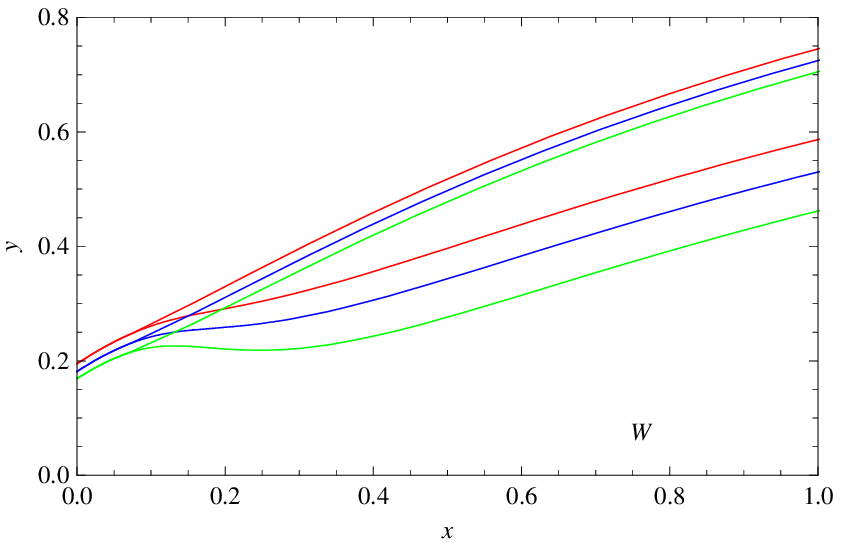}
&
  \psfrag{y}[][]{\scriptsize{$\sqrt{\Lambda_-/3}~\bar r$}}
  \psfrag{x}[][]{\scriptsize{$\sqrt{\Lambda_-/3}~t$}}
  \psfrag{W}[][]{\scriptsize{\fbox{$w=-1 \rightarrow w=1/3$}}}
  \includegraphics[width=0.438\textwidth]{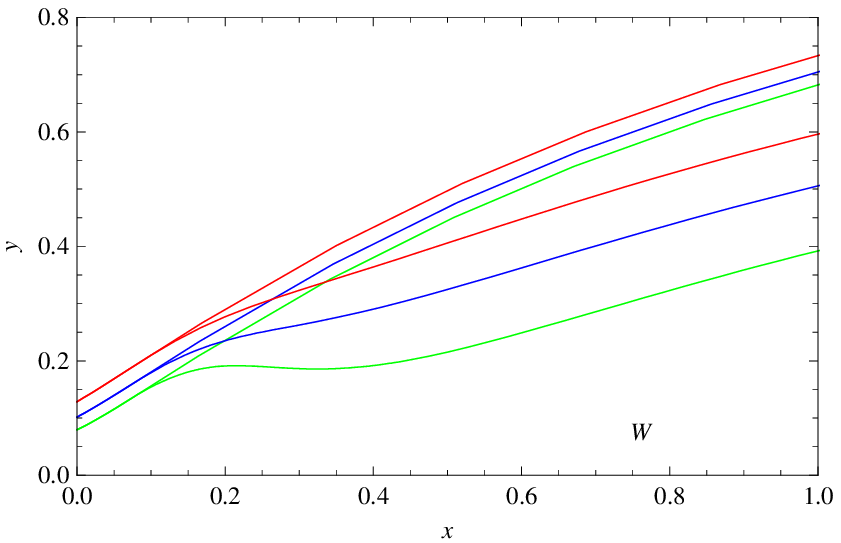}\\

  \psfrag{y}[][]{\scriptsize{$\sigma/\sigma_0$}}
  \psfrag{x}[][]{\scriptsize{$\sqrt{\Lambda_-/3}~t$}}
  \psfrag{W}[][]{\scriptsize{\fbox{$w=1/3 \rightarrow w=-1$}}}
  \includegraphics[width=0.438\textwidth]{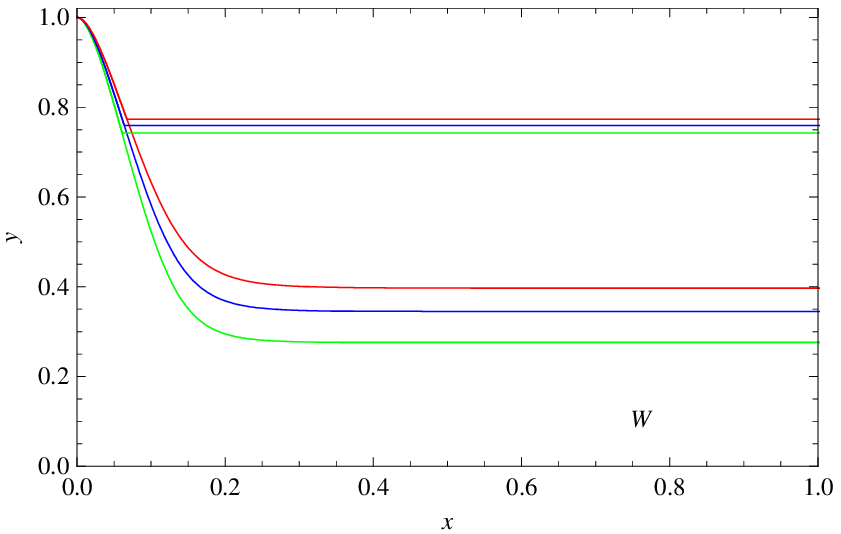}
&
  \psfrag{y}[][]{\scriptsize{$\sigma/\sigma_0$}}
  \psfrag{x}[][]{\scriptsize{$\sqrt{\Lambda_-/3}~t$}}
  \psfrag{W}[][]{\scriptsize{\fbox{$w=-1 \rightarrow w=1/3$}}}
  \includegraphics[width=0.438\textwidth]{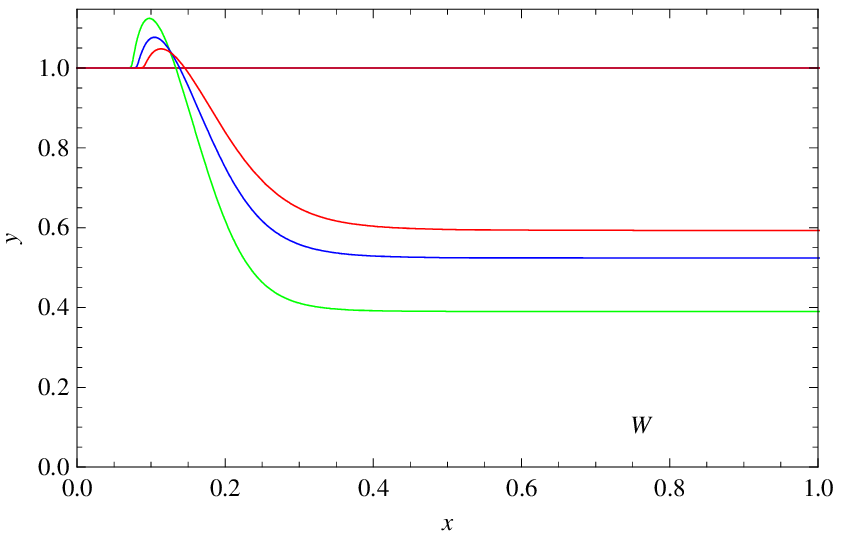}
\end{tabular}
  \caption{\label{ambiPTfig}
Numerical survey of the propagation of a vacuum bubble in an environment (FRW) concurrently undergoing a reheating PT or vice versa. Three different initial values $\sigma_0$ (red, blue, green) are indicated respectively. The \emph{upper row} shows the resulting bubble trajectories in the FRW frame: while in the reheating case the expansion of the bubble is reversed by the passage through the ambient PT, in the $w=1/3 \rightarrow w=-1$ case the contraction of the bubble is stopped by the transition. Different to the LTB case notable effects remain to be seen also in the interior frame (\emph{middle row}). Analogously to the LTB case a pronounced effect on the surface tension itself is found (\emph{lower row}) which is a consequence of the jointly used junction method.}
\end{center}
\end{figure}

Another interesting dynamical background is a FRW model containing a perfect fluid that undergoes a rapid PT (e.g.~reheating) while the vacuum bubble propagates through it. A numerical study confirms that the PT of the background does change the trajectory of the bubble notably and that, unlike in the previous case, this can also be seen from the interior, see Fig.~\ref{ambiPTfig}.


\section{Conclusions}

Rapid tunneling processes on the landscape may give rise to the nucleation of bubbles of new vacuum on a diversity of backgrounds. This kind of first-order PTs can be relevant for cosmological scenarios, like e.g.~chain inflation. Therefore the nucleation as well as the evolution of de Sitter bubbles, which can be toy models of inflation, on any non-standard backgrounds is worth studying.

Concerning the nucleation of vacuum bubbles we have analysed the possible effect on the tunneling rate with an explicitly time dependent power law inflating FRW backround. We found that the rate can indeed be enhanced w.r.t.~the quasistatic or Minkowskian approximations, if the crossing time scale of nucleated bubbles is larger than the time scales characterising the background dynamics. We have also seen that in a radiation dominated background the particle horizon constrains the nucleation scale of bubbles and that bubbles nucleated with about horizon scale become responsive of details of the cosmology close to the Big Bang.

In order to study the classical evolution of a de Sitter bubble on non-standard backgrounds we used Israel junction conditions to model bubble propagation in a matter environment. We saw already in the homogeneous limit that dust dominated backgrounds prevent bubbles from growing. Nevertheless, for the inhomogeneous LTB background, initially expanding bubbles can be found. We numerically studied the effect of ambient inhomogeneity on such bubbles and found that possible disturbances of the wall trajectory would not be seen by inside observers. We also analysed the effects of a rapid PT in a FRW background (e.g.~reheating). Our analysis suggests that in this scenario perturbations of the bubble wall are present also for observers inside the bubble.

Within the context of primordial bubble collisions it has been pointed out \cite{Aguirre08,Chang08} that similar disturbances of the bubble trajectory could in principle be observable in the CMB.

\end{document}